\documentclass[preprint,preprintnumbers,amsmath,amssymb]{revtex4}

\usepackage{epsfig}
\usepackage{amsfonts}

\usepackage{amssymb}
\begin{document}


\title{Superstatistical Brownian motion}

\author{Christian Beck}
\affiliation{School of Mathematical Sciences, Queen Mary,
University of London, Mile End Road, London E1 4NS, UK}

\date{\today}

\vspace*{2cm}

\begin{abstract}
As a main example for the superstatistics approach, we study a
Brownian particle moving in a $d$-dimensional inhomogeneous
environment with macroscopic temperature fluctuations. We discuss
the average occupation time of the particle in spatial cells
with a given temperature. The Fokker-Planck equation for
this problem becomes a stochastic partial differential equation.
We illustrate our results using experimentally measured time
series from hydrodynamic turbulence.
\end{abstract}

\maketitle

\section{Introduction}

Nonextensive statistical mechanics \cite{tsa1,tsa2,tsa3,abe}, 
originally developed as an
equilibrium formalism, has indeed many interesting applications
for driven nonequilibrium systems of sufficient complexity.
Consider e.g.\ a Brownian particle moving through a changing
environment. We may assume that the environment is inhomogeneous
and a suitable parameter describing the state of this
environment, e.g.\ the inverse temperature $\beta$, fluctuates on
a relatively large spatio-temporal scale. This means that there is a
fast dynamics given by the velocity of the Brownian particle and
a slow one given by temperature changes of the environment. The
two effects produce a superposition of two statistics, or in a
short, a superstatistics \cite{beck-cohen, boltzmann-m,
beck-su, touchette-beck, yamano, plastino,
ryazanov, prl, wilk, souza, souza2, luczka, garcia, sattin,
sattin2, erice, chavanis, haenggi, grigolini}.
The stationary probability
distributions for an ordinary Brownian particle in a constant
environment are certainly Gaussian distributions, but for a fluctuating
environment one obtains non-Gaussian behaviour with fat tails.
These tails can decay e.g. with a power law, or as a stretched
exponential, or in an even more complicated way
\cite{touchette-beck}. Which type of tails are produced depends on
the probability distribution $f(\beta)$ of the parameter $\beta$.
For the above simple example of a generalized Brownian particle,
$\beta$ is the fluctuating inverse temperature of the
environment, but in general $\beta$ can also be an effective
friction constant, a changing mass parameter, a changing
amplitude of Gaussian white noise, the fluctuating energy
dissipation in turbulent flows, or simply a local variance
parameter extracted from a signal. Recent applications of
the superstatistics concept include a variety of physical
systems, such as Lagrangian \cite{reynolds, beck03, boden, aringazin} and
Eulerian turbulence \cite{beck-physica-d, jung-swinney, BCS}, defect
turbulence \cite{daniels}, atmospheric turbulence
\cite{rapisarda, rap2}, cosmic ray statistics \cite{cosmic}, solar
flares \cite{maya}, solar wind statistics
\cite{burlaga}, networks \cite{abe-turner, hasegawa}, random
matrix theory \cite{abul-magd}, and mathematical finance \cite{bouchard,
ausloos, hasegawa2}.

If $\beta$ is distributed according to a particular probability
distribution, the $\chi^2$-distribution, then the corresponding
marginal stationary velocity distributions of the Brownian particle
obtained by integrating over all $\beta$ are given by the
generalized canonical distributions of nonextensive statistical
mechanics \cite{tsa1,tsa2,tsa3,abe}. For
other distributions of the intensive parameter $\beta$, one ends
up with more general statistics, which contain Tsallis statistics
as a special case.

In this paper we first briefly review the superstatistics concept.
We then look at typical occupation times of the superstatistical
Brownian particle in cells of constant temperature, 
distinguishing between temporal and spatial temperature fluctuations.
In section 4 we emphasize a certain analogy between
the step from statistics to superstatistics and the step from
1st quantization to 2nd quantization:
In both cases the rate
equations for probability densities are upgraded from
deterministic partial differential equations to stochastic ones.
Finally, in the last section we compare with experimental data in
turbulent flows, where superstatistical models are a very useful
tool to effectively describe the dynamics.

\section{Various types of superstatistics}

It is well known that for equilibrium systems of ordinary
statistical mechanics the probability to observe a state with energy $E$
is given by
\begin{equation}
p(E)=\frac{1}{Z(\beta)}\rho(E) e^{-\beta E}.
\end{equation}
This formula just describes the ordinary canonical ensemble.
$e^{-\beta E}$ is the Boltzmann factor, $\rho(E)$ is the density
of states and $Z(\beta)$ is the normalization constant of $\rho
(E)e^{-\beta E}$. For superstatistical systems, one generalizes
this approach by assuming that $\beta$ is a random variable as
well. Indeed, a driven nonequilibrium system is often
inhomogeneous and consist of many spatial cells with different
values of $\beta$ in each cell. 
The cell size is effectively
determined by the correlation length of the continuously varying
$\beta$-field. If we assume that each cell reaches local
equilibrium very fast, i.e.\ the associated relaxation time is
short, then in the long-term run the stationary probability
distributions arise as a superposition
of Boltzmann factors $e^{-\beta E}$ weighted with the probability
density $f(\beta)$ to observe some value $\beta$ in a randomly
chosen cell:
\begin{equation}
p(E)=\int_0^\infty f(\beta)  \frac{1}{Z(\beta)} \rho(E) e^{-\beta
E}d\beta  \label{ppp}
\end{equation}

The simplest example is a Brownian particle of mass $m$ moving
through a changing environment in $d$ dimensions. For its
velocity $\vec{v}$ one has the local Langevin equation
\begin{equation}
\dot{\vec{v}}=-\gamma \vec{v} + \sigma \vec{L}(t) \label{lange}
\end{equation}
($\vec{L}(t)$: $d$-dimensional Gaussian white noise, $\gamma$:
friction constant, $\sigma$: strength of noise) which becomes
superstatistical because for a fluctuating environment the
parameter $\beta :=\frac{2}{m} \frac{\gamma}{\sigma^2}$ becomes a
random variable as well: it varies from cell to cell on the large
spatio-temporal scale $T$. Of course, for this example
$E=\frac{1}{2}m\vec{v}^2$. During the time scale $T$ the local
stationary distribution in each cell is Gaussian,
\begin{equation}
p(\vec{v}|\beta)=\left( \frac{\beta}{2\pi}\right)^{d/2}
e^{-\frac{1}{2}\beta m\vec{v}^2}.
\end{equation}
But the marginal distribution describing the long-time behavior of
the particle for $t>>T$,
\begin{equation}
p(\vec{v})=\int_0^\infty f(\beta)p(\vec{v}|\beta)d\beta
\label{margi}
\end{equation}
exhibits fat tails. The large-$|v|$ tails of the distribution
(\ref{margi}) depend on the behaviour of $f(\beta )$ for $\beta
\to 0$ \cite{touchette-beck}. Many different superstatistical
models corresponding to different $f(\beta)$ are possible: The
function $f$ is determined by the environmental dynamics of the
nonequilibrium system under consideration. 

Consider, for example,
a simple case where there are $n$ independent Gaussian random variables $X_1,
\ldots , X_n$ underlying the dynamics of $\beta$ in an additive
way.
$\beta$ needs to be positive and
a positive $\beta$ is obtained by squaring
these Gaussian random variables. The probability distribution
of a random variable that is the sum of
squared Gaussian random variables
$\beta=\sum_{i=1}^nX_i^2$ is well known in statistics:
It is the $\chi^2$-distribution of degree $n$,
i.e. the probability density $f(\beta)$ is given by
\begin{equation}
f(\beta )=\frac 1{\Gamma (\frac n2)}\left( \frac n{2\beta _0}\right)
^{n/2}\beta ^{n/2-1}e^{-\frac{n\beta }{2\beta _0}},  \label{chi2}
\end{equation}
where $\beta_0$ is the average of $\beta$.

One can now simply do the integration in eq.~(\ref{margi}) using
eq.~(\ref{chi2}) \cite{prl, wilk}. The result is the generalized
canonical distribution of nonextensive
statistical mechanics, i.e.\ a $q$-exponential of the form
\begin{equation}
p(\vec{v}) \sim \frac{1}{(1+\tilde{\beta}
(q-1)\frac{1}{2}m\vec{v}^2)^{\frac{1}{q-1}}}
\end{equation}
with
\begin{equation}
q=1+\frac{2}{n+d}
\end{equation}
and
\begin{equation}
\tilde{\beta}=\frac{2\beta_0}{2-(q-1)d}.
\end{equation}
Thus these types of generalized Boltzmann factors
occur as stationary states for nonequilibrium systems with
suitable fluctuations of some intensive parameter $\beta$.

Instead of $\beta$ being a sum of many contributions, for other
systems the random variable $\beta$ may be generated by
multiplicative random processes. In this case one typically ends
up with a log-normally distributed $\beta$ \cite{BCS}, i.e.\  the probability
density is given by
\begin{equation}
f(\beta )=\frac{1}{\sqrt{2\pi}s\beta}
\exp \left\{ \frac{-(\ln \frac{\beta}{\mu})^2}{2s^2}\right\},
\label{logno}
\end{equation}
where $\mu$ and $s$ are parameters. Both $\chi^2$-superstatistics
($=$ Tsallis statistics) and lognormal superstatistics are
relevant for turbulent systems. The observation is that the
former one particularly well describes atmospheric turbulence
data \cite{rapisarda, rap2}, whereas the latter one gives good
results for laboratory turbulence (e.g.\ Taylor-Couette flow)
\cite{beck-physica-d, jung-swinney, BCS}. Note that the Reynolds number is
controlled for laboratory turbulence, whereas it fluctuates for
the atmospheric case.

\section{Occupation times in cells of constant local temperature}

Let us now discuss some subtleties of the superstatistical
modeling approach. An important question is what really causes the
fluctuations of temperature around the Brownian particle. Clearly
these are produced by the fact that we consider a driven
nonequilibrium system in a stationary state which has some
external energy input and also some energy dissipation (like the sun
acting on the earth's atmosphere and producing spatio-temporal
temperature fluctuations in form of the weather). There are,
however, different possibilities:
Either there could be explicit temporal temperature
changes around the Brownian particle, or the particle could move through
spatial cells of size $L^d$ which all have different temperatures
but no explicit time dependence.
The latter case is more a spatial then a temporal effect, and corresponds
to a frozen random pattern.

To treat the second case properly, we have to take into account
the average occupation time of the Brownian particle in a cell of
size $L^d$ where the temperature is approximately constant.
Locally, in each cell the velocity of our particle is described by
the Ornstein-Uhlenbeck process and its position by the
Ornstein-Uhlenbeck position process \cite{vKa}. The latter process
is a Gaussian diffusion process. For each position component
$x_i(t)$, $i=1, \ldots , d$ of the Brownian particle one has for
large times $t$
\begin{equation}
\langle x_i^2 (t) \rangle =2Dt,
\end{equation}
where the diffusion constant $D$ is given by the Einstein relation
\begin{equation}
D= \frac{1}{m\gamma \beta}.
\end{equation}
Of course, the larger the temperature $\beta^{-1}$ in a given
cell, the
faster the particle will diffuse, hence it will occupy a given
cell of size $L^d$ for a shorter time on average if the
temperature is higher in that cell. In fact, the average time $\bar{t}$
the particle spends in a cell of temperature $\beta^{-1}$
is given by
\begin{equation}
L^2=2D\bar{t} \Longleftrightarrow
\bar{t}=\frac{L^2}{2D}=\frac{1}{2}L^2m\gamma \beta .
\end{equation}
Assuming that on average all spatial cells with a given local
temperature have the same typical size $L^d$, this means the
average occupation time $\bar{t}$ in a given cell is proportional to
$\beta$. The probability to find the particle in a given cell is
proportional to the occupation time, hence this means we have to
adjust the probability density $f_{space}(\beta)$ describing the
spatial distribution of $\beta$ in the various cells by a factor $\beta$
if we proceed to a temporal description $f_{temp}(\beta)$, i.e.\ if we take into
account how much time the particle spends in each cell:
\begin{equation}
f_{temp}(\beta) \sim \bar{t} f_{space} (\beta ) \sim \beta f_{space} (\beta)
\end{equation}
In general, 
one has for a given
effective energy $E$ the local Boltzmann-Gibbs distributions
\begin{equation}
p_{loc}(E|\beta )=\beta e^{-\beta E}
\end{equation}
in each cell (assuming for simplicity a constant density of states).
These distributions are properly normalized,
\begin{equation}
\int_0^\infty p_{loc} (E|\beta )dE= \int_0^\infty \beta e^{-\beta E}dE=1.
\end{equation}
Taking temporal averages over all
possible $\beta$, one obtains the marginal distributions
\begin{equation}
p(E)=\int_0^\infty \beta f_{temp}(\beta) e^{-\beta E}d\beta .
\end{equation}
Our previous consideration shows that proceeding to a
spatial probability density $f_{space} (\beta)$ describing the
distribution of $\beta$ in the various spatial cells requires 
an additional weighting by a factor $\beta$, since
a typical particle will spend more time in cells with small
temperature. 
We thus arrive at
\begin{equation}
p(E) \sim  \int_0^\infty \beta^2 f_{space} (\beta) e^{-\beta
E}d\beta .
\end{equation}
Our consideration shows that care has to be taken
in the interpretation
of the probability density $f(\beta)$, namely whether it
describes the distribution of $\beta$ in spatial cells or in
temporal time slices. In general, one must also take into account
the density of state and the fact that for Brownian particles
probability densities are usually regarded as a function of the
velocity $\vec{v}$ rather than the energy $E=\frac{1}{2}m\vec{v}^2$. This
corresponds to a simple transformation of random variables.

\section{Stochastic Fokker-Planck equation}

Let us once again emphasize that for a superstatistical Brownian
particle one has a superposition of two stochastic processes on
different time scales: A fast stochastic process given by a local
Ornstein-Uhlenbeck process and a slow stochastic process given by
$\beta (t)$. The $\beta$-process can a priori be anything. We assume that
the parameters $\sigma$ and $\gamma$ in eq.~(\ref{lange}) are
constant for a sufficiently long time scale $T$, and then change
to new values, either by an explicit time dependence, or by a
change of the environment through which the Brownian particle
moves.

For simplicity, let us consider the 1-dimensional case.
Locally, in a given cell with
constant temperature the probability density $P(v,t)$ obeys the Fokker-Planck
equation \cite{vKa}
\begin{equation}
\frac{\partial P}{\partial t}=\gamma \frac{\partial (vP)}{\partial
v}+ \frac{1}{2} \sigma^2 \frac{\partial^2 P}{\partial v^2}
\end{equation}
with the local stationary solution
\begin{equation}
P(v|\beta)=\sqrt{\frac{m\beta}{2\pi}} \exp \left\{-\frac{1}{2}
\beta mv^2 \right\} \label{28}.
\end{equation}
In the superstatistical
approach, one regards $\gamma (t)$ and
$\sigma (t)$ as stochastic processes as well, which
evolve on a much larger typical time scale $T$ as
compared to $v(t)$. This means that for
superstatistical systems the above equation is not an ordinary
Fokker-Planck equation: It is a Fokker-Planck equation whose
coefficients $\gamma (t)$ and $\sigma (t)$ are stochastic
processes, with all their properties to be specified
by a characteristic functional. Hence, in
the superstatistical approach the equation that determines the
evolution of probabilities becomes a stochastic partial
differential equation---rather than a deterministic one.
At the end, one has to perform an average over all realizations
of $\gamma (t)$ and $\sigma (t)$. Needless to say that one can
generalize all this to superstatistical versions of higher-dimensional
and nonlinear Fokker-Planck
equations as well.

We can see a certain analogy to the process of second quantization
in quantum field theories. Consider, for example, a Klein-Gordon
(or Schr\"odinger) field. The Klein-Gordon (or
Schr\"odinger) equation governs the time evolution of a
wave function, and the wave function squared essentially gives the
quantum mechanical probability density in a first-quantized system. 
Thus the Klein-Gordon (or Schr\"odinger) equation corresponds to
a deterministic evolution equation for
probability densities. In our case, this role is taken over by
the local (deterministic) Fokker-Planck equation. In order to
proceed to second quantization in quantum mechanics a possible
method is to add noise to the classical field equation (e.g. the
Klein-Gordon equation), obtaining a second-quantized theory via
the stochastic quantization procedure \cite{parisi-wu, dam}. By
second quantization, we thus construct a stochastic partial
differential equation, a Klein-Gordon equation
with noise. In a similar way, for superstatistical
systems we proceed from a deterministic Fokker-Planck equation to
a stochastic one, by replacing the constant parameters $\gamma$ or
$\sigma$ by suitable stochastic processes. Thus proceeding to a
superstatistical description is analogous to proceeding to a
second quantized description in quantum mechanics. 

To completely
solve the superstatistical system dynamics one has to find all solutions
of the stochastic Fokker-Planck equation (which are stochastic
processes) and then form expectations. In full generality, this
is quite a heavy task, though some progress has been made for
very simple dichotomous models \cite{luczka}.
The solution of the
stochastic Fokker-Planck equation 
depends on what type of
stochastic processes are chosen for $\gamma (t)$ and
$\sigma (t)$. 
As shown in \cite{erice}, one
has for example a power-law decay of the marginal velocity correlation
function if $\gamma$ fluctuates and $\sigma$ is constant, and an
exponential decay if $\gamma$ is constant and $\sigma$
fluctuates, choosing the same $\chi^2$-distribution of $\beta
=\frac{2}{m}\frac{\gamma}{\sigma^2}$ in both cases and assuming
$T>>\tau$, $\tau$ being the local relaxation time
in each cell. To construct the best
superstatistical model for a given physical application, one has
to carefully compare with the experimental data, looking at both,
stationary densities and correlation functions.

\section{Comparison with turbulent velocity fluctuations}

A major application of superstatistics is in turbulence
modeling \cite{prl, beck03, reynolds, boden, aringazin, beck-physica-d,
jung-swinney, BCS}.
Here essentially the local
velocity $v$ of the superstatistical Brownian particle model 
corresponds to a local
longitudinal velocity difference $u$ in the turbulent flow, and $\beta$ is a
local variance parameter connected with fluctuations in energy
dissipation. As said before, to construct a suitable
superstatistical turbulence model one has to compare carefully
with various experimentally accessible quantities. 
For the example of a measured velocity time series in a
turbulent Taylor-Couette flow, one basically observes the
following \cite{BCS}: The probability density $f_{temp}(\beta)$ 
is essentially a
lognormal distribution (Fig.~1).
\begin{figure}
\epsfig{file=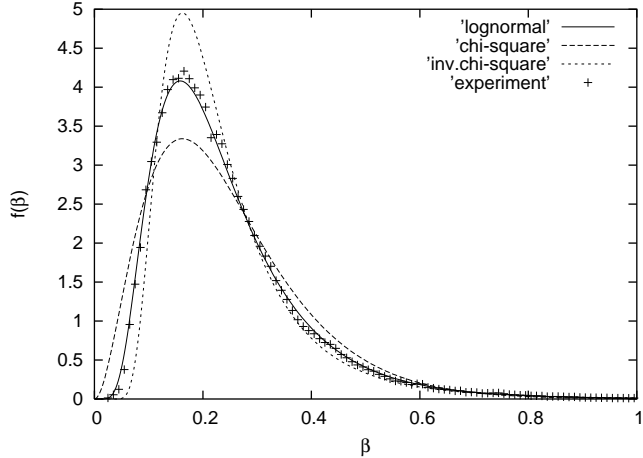} \caption{Probability density
$f_{temp}(\beta)$ as extracted from an experimentally measured time
series in a turbulent Taylor-Couette flow \cite{BCS}. Also shown is
a lognormal, $\chi^2$, and inverse $\chi^2$ distribution with the
same mean and variance as the experimental data.}
\end{figure}
Correlation functions of the measured velocity signal $v(t)$ decay
more or less exponentially, but those of temporal velocity differences
$u(t)=v(t+\delta)-v(t)$ on a a given time scale $\delta$ decay
asymptotically with a power law (Fig.~2). The
marginal distribution $p(u)$ exhibits fat
tails whose flatness decreases with increasing $\delta$. On all
scales the measured histograms
$p(u)$ are well described by lognormal superstatistics
as given by eq.~(\ref{margi}) and (\ref{logno}), 
replacing $\vec{v}\to u$. One observes a
\begin{figure}
\epsfig{file=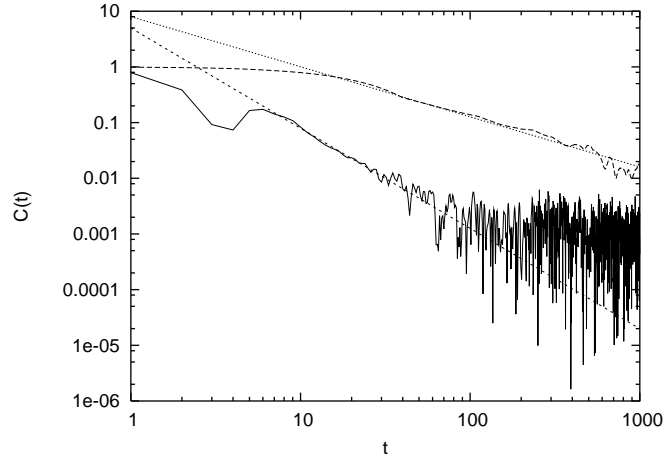} \caption{Correlation functions
$C_\beta (t)$ (top) and $|C_u(t)|$ (bottom) as measured in
a turbulent Taylor-Couette flow \cite{BCS} on the smallest scales.
The straight lines show power laws with exponents
-0.9 and -1.8.}
\end{figure}
a clear time-scale separation between the 
process $u(t)$ and the corresponding process $\beta (t)$:
$\beta (t)$ evolves on a much larger typical time scale. This
time scale separation increases with increasing Reynolds number,
as well as with $\delta$ \cite{BCS}. The process $\beta (t)$ is
strongly correlated, its correlation function decays with a power
law.

\end{document}